\documentclass[prl,twocolumn,superscriptaddress]{revtex4}
\usepackage{graphicx}
\usepackage{subfigure}
\usepackage{amsmath}
\usepackage{braket}
\usepackage{amssymb}

\usepackage{fancyhdr}

\newcommand{\be}{\begin{equation}}
\newcommand{\ee}{\end{equation}}
\newcommand{\bea}{\begin{eqnarray}}
\newcommand{\eea}{\end{eqnarray}}
\newcommand{\SKIP}[1]{}

\begin{document}



\title{Observing complex bound states in the spin-1/2 Heisenberg XXZ
  chain using local quantum quenches}
\author{Martin Ganahl}
\affiliation{Institut f\"ur Theoretische Physik, Technische
  Universit\"at Graz, Petergasse 16, 8010 Graz, Austria}
\author{Elias Rabel}
\affiliation{Institut f\"ur Theoretische Physik, Technische
  Universit\"at Graz, Petergasse 16, 8010 Graz, Austria}
\affiliation{Institut f\"ur Festk\"orperforschung, 
  Forschungszentrum  J\"ulich,  52425  J¨\"ulich,  Germany}
\author{Fabian H.L. Essler}
\affiliation{The Rudolf Peierls Centre for Theoretical Physics, 
  Oxford University, Oxford OX1 3NP, UK.}
\author{H.G. Evertz}
\email{evertz@tugraz.at}
\affiliation{Institut f\"ur Theoretische Physik, Technische
  Universit\"at Graz, Petergasse 16, 8010 Graz, Austria}

\date{December 14, 2011} 

\pagestyle{fancy}
\fancyhf{}
\fancyhead[L]{Ganahl, Rabel, Essler, Evertz}
\fancyhead[R]{Observing complex bound states 
}

\begin{abstract}
We consider the non-equilibrium evolution in the spin-1/2 XXZ
Heisenberg chain for fixed magnetization after a \emph{local quantum
quench}. This model is equivalent to interacting spinless fermions.
Initially an infinite magnetic field is applied to $n$
consecutive sites and the ground state is calculated. At time $t=0$
the field is switched off and the time evolution of observables such
as the z-component of spin is computed using the Time Evolving Block
Decimation (TEBD) algorithm. We find that the observables exhibit
strong signatures of linearly propagating spinon and bound state
excitations. These persist even when integrability-breaking
perturbations are included. Since bound states (``strings'') are
notoriously difficult to observe using conventional probes such as
inelastic neutron scattering, we conclude that local quantum quenches
are an ideal setting for studying their properties. We comment on
implications of our results for cold atom experiments. 

\end{abstract}


\maketitle

Cold atomic gases provide an ideal testing ground for non-equilibrium
many-body quantum physics because the dynamics remains coherent for
long times by virtue of the weak coupling to the environment. 
Recent experiments \cite{uc,kww-06}
have opened up the study of an entirely new regime in many
particle quantum physics. The ``quantum Newton's cradle''
experiments of Kinoshita et al drew attention to the importance of
dimensionality and conservation laws and prompted a huge number of
theoretical analyses on the role played by quantum integrability
\cite{reviews,integrability}.
A standard protocol for driving a quantum system out of equilibrium is
by means of a quantum quench (QQ): a system is prepared in the ground state of
a given Hamiltonian $H_0$. At time $t=0$ an experimentally tuneable
parameter that characterizes the Hamiltonian (e.g. a magnetic field)
is changed suddenly and one then considers the unitary time evolution of
the system by means of the new Hamiltonian $H$. QQs can be either
global or local 
and we focus on the latter case in the
following. A particular case of a local QQ is given by the X-ray edge
singularity, which is a central paradigm of many-body physics. The
types of problems we consider below can be viewed as generalizations of
X-ray edge problems, the most crucial difference arising from the initial state and the kind of
observable that we consider, which can be measured e.g. in
realizations based on cold atomic gases.

We consider the anisotropic spin-1/2 Heisenberg chain on a lattice
with $N$ sites with fixed numbers $N_{\uparrow,\downarrow}$ of up and
down spins and open boundary conditions
\footnote{The boundary conditions do not affect the propagation
  noticeably until the wave front reaches the boundary.})
\begin{align}
H(\Delta,B_0) &= J\sum_{i=1}^{N-1}S^x_iS^x_{i+1}+S^y_iS^y_{i+1}+
\Delta S^z_iS^z_{i+1}\nonumber\\
&-B_0(t)\sum_{i=i_0}^{i_0+m_0-1}S^z_i,
\label{eq:Heisenberg hamiltonian}
\end{align}
where $J>0$ and $B_0$ is a local magnetic field acting on
$m_0$ consecutive sites starting at position $i_0$. It is well-known that
(\ref{eq:Heisenberg hamiltonian}) can be mapped to a model of spinless
fermions with nearest-neighbour density-density interaction by means
of a Jordan-Wigner transformation and all of our results are
straightforwardly translated into that setting. 
The study of local QQs in models of the kind (\ref{eq:Heisenberg
  hamiltonian}) was initiated in 1970 \cite{doug}, where the
noninteracting case $\Delta=0$, $m_0=1$ was shown to lead to a
non-thermal stationary state.
With the advent of efficient numerical approaches \cite{tebd,tDMRG},
local quenches in the interacting XXZ chain 
\cite{XXZ_local_quenches,Pereira_2008}
and corresponding conformal field theories \cite{LQQ} have been studied intensely.
In the present letter we show that longer quenches $m_0>1$ lead to 
prominent linearly propagating bound states,
which in standard condensed matter scenarios have been 
difficult to discern \cite{Pereira_2008,structurefactor}.

We consider the following quench protocol: we prepare the system
in the ground state $|0\rangle$ of the Hamiltonian
$H(\Delta,B_0=-\infty)$. At time $t=0$ we suddenly switch off the
magnetic field $B_0$ and then consider the time evolution, 
governed by the Hamiltonian $H(\Delta,B_0=0)$, of the following
observables 
\bea
\langle S^z \rangle(j,t)&\equiv&\langle 0|S^z_j(t)| 0\rangle\
,\nonumber\\
P_{\uparrow\uparrow}(j,t)&\equiv&\langle 0|
P_j(t)P_{j+1}(t)
| 0\rangle\ ,\nonumber\\
P_{\uparrow\uparrow\uparrow}(j,t)&\equiv&\langle 0|
P_{j-1}(t)P_j(t)P_{j+1}(t)
| 0\rangle\ ,
\label{observables}
\eea
where $P_j=S^z_j+1/2$ is the up-spin projection operator on site $j$.
In the thermodynamic limit there are different regimes: when
the magnetization per site $m$ is equal to $-1/2$ the ground state of
$H(\Delta,B_0=0)$ is given by the saturated ferromagnetic state with
all spins down and a local quench of the type described above then
reduces to a quantum mechanical few-body problem. On the other hand,
for magnetizations $-1/2<m<0$ the model $H(\Delta,B_0=0)$ describes a
quantum critical (Luttinger liquid) phase and our local quantum quench
involves complex many-body effects and can be thought of as a
generalization of the X-ray edge problem. In the following we first
consider the simpler, spin-polarized case as this allows us to
establish the role played by bound states.

\emph{Spin Polarized Case:} In this case the ground
state of $H(\Delta,0)$ is the ferromagnetic state with all spins
down $|\downarrow\rangle$. Excitations with $N_\uparrow$ 
spin-flips (particles) can be constructed by Bethe's Ansatz and are
parametrized by $N_\uparrow$ momenta $k_j$ 
\bea
|N_\uparrow,\bf{k}\rangle&=&\!\!\!\!
\sum_{x_1<\dots<x_{N_\uparrow}}\!\!\!\!\Psi\big(\{k_j\}|\{x_l\}\big)
\prod_{n=1}^{N_\uparrow} S^+_{x_n}|\downarrow\rangle.
\eea
Here the wave function $\Psi$ has the characteristic Bethe Ansatz form
and the momenta $\{k_j\}$ are subject to quantization conditions,
which for a ring geometry read
\be
e^{iNk_j}=\prod_{l=1\atop l\neq j}^{N_\uparrow}
-\frac{2\Delta e^{ik_j}-1-e^{ik_j+ik_l}}
{2\Delta e^{ik_l}-1-e^{ik_j+ik_l}}
\ ,\ j=1,\dots,N_\uparrow.
\label{bae}
\ee
Energy and momentum are $E=\sum_{j=1}^{N_\uparrow}\epsilon(k_j)$ and
$P=\sum_{j=1}^{N_\uparrow}k_j$ respectively, where $\epsilon(k)=
J\big(\cos k-\Delta\big)$. The solutions $k_j$ of 
(\ref{bae}) can be either real or complex \cite{strings}. The former
describe scattering states of ``magnons'', while the latter correspond
to bound states. Bound states involving $\ell$ particles are known as
``$\ell$-strings" and have wave functions that exhibit exponential
decay (which can be slow) with respect to the distances between
particles. Their dispersion relations in
the thermodynamic limit are \cite{strings,sutherland_beautiful_models}
$\epsilon_\ell(k)=-J\frac{\sin(\nu)}{\sin(\ell\nu)}\big(\cos(\ell\nu)-
(-1)^\ell\cos(k)\big)$, where $\Delta = \cos(\nu)$. Here the total momentum
$k$ of $\ell$-strings is constrained, e.g.
for $|\Delta |<1$ and $\ell=2$ we have $|k|>2\nu$.
For a given value of $\Delta$ there generally exists a hierarchy of
allowed strings, which was first identified in a seminal work by
Suzuki and Takahashi \cite{strings}. We note that the energy
difference between bound states and scattering continua can generally be
very small. Using the exact eigenstates of $H(\Delta,0)$ we can derive
a Lehmann representation for the observables (\ref{observables}) after
our quench 
\bea
\langle {\cal O}\rangle(j,t)\!&=&\!\sum_{\{k_l\},\{p_r\}}
\langle 0|m_0,{\bf k}\rangle
\langle m_0,{\bf k}|{\cal O}_1|m_0,{\bf p}\rangle
\langle m_0,{\bf p}|0\rangle
\nonumber\\
&\times&
e^{-i\sum_{n=1}^{m_0}t[\epsilon(p_n)-\epsilon(k_n)]-(j-1)[p_n-k_n]]}\ ,
\label{lehmann}
\eea
where the sums are over all Bethe Ansatz states with $m_0$ momenta. 
In the case $m_0=1$ an elementary calculation gives $\langle
S^z\rangle(j,t)=-\frac{1}{2}+J_{j-1}^2(Jt)$, where $J_n$ is a Bessel
function. For large, fixed $j$ this increases exponentially for 
$Jt\alt j$, shows a maximum for $Jt\approx j$ and exhibits an
oscillatory power-law decay for $Jt\agt j$. 
A stationary phase approximation shows that the dominant contribution
in the Lehmann representation (\ref{lehmann}) for $Jt\approx j$ arises
from states with $k\approx\frac{\pi}{2},\frac{3\pi}{2}$, which
propagate with the highest possible velocity $v_{\rm max}={\rm
  max}_k\big|\frac{\epsilon(k)}{dk}\big|=J$. The fact that
$\langle S^z\rangle(j,t)$ has a maximum at $Jt\approx j$ can be
understood qualitatively by noting that the density
of states (DOS) $\rho_1(v) = \int \delta(v-d\epsilon/dk)dk =
\frac{N}{2\pi}\frac{1}{\sqrt{J^2-v^2}}$ has singularities at the maximum
speed $v=\pm J$. The exponential supression of
$\langle S^z\rangle(j,t)$ for $t\alt (j/v_{\rm max})$ gives rise
to a horizon effect and is described by the Lieb-Robinson bound \cite{lr}.

\begin{figure}
  \includegraphics[width=9cm,height=8cm]{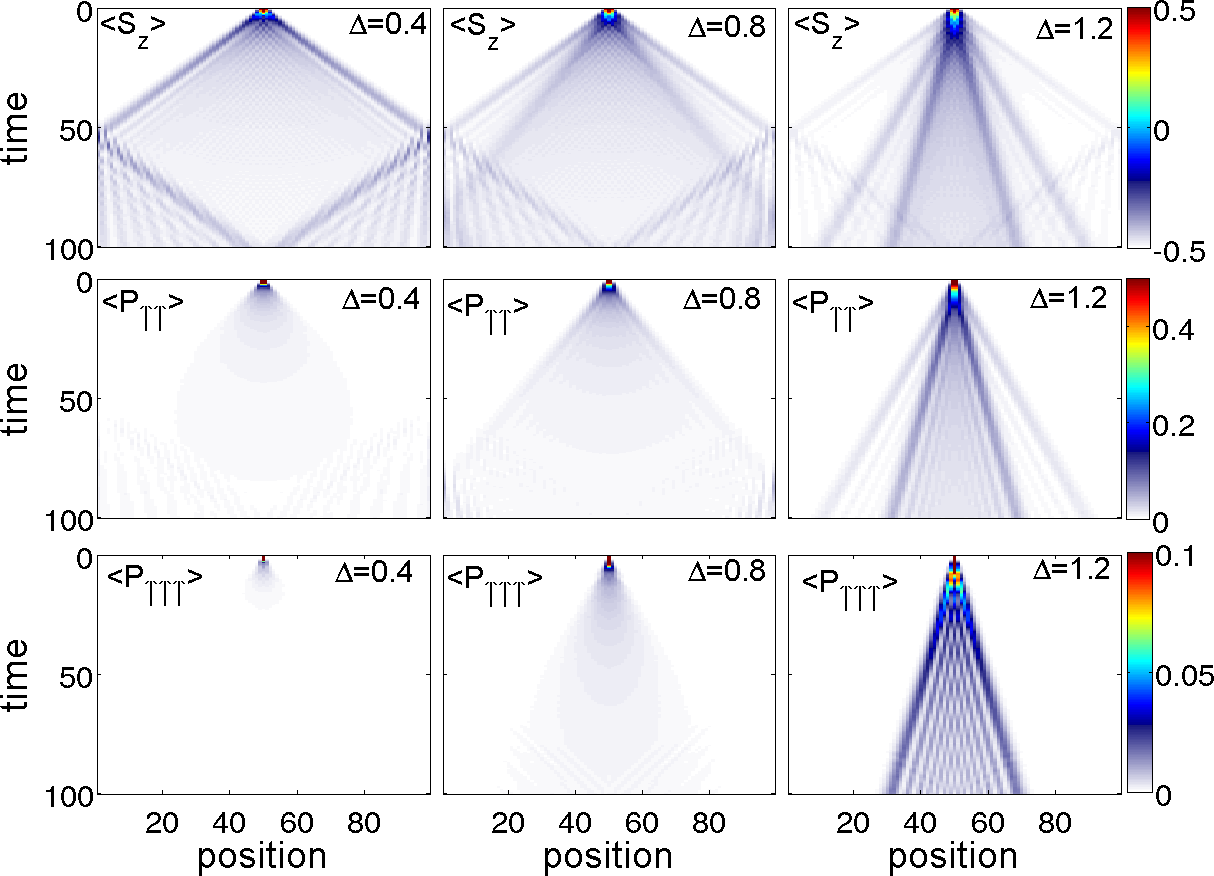}
\caption{Time evolution in the spin polarized case after preparing the
system in a initial state with three spin flips in the centre of a 101
site chain for different values of $\Delta$. Top row:
Spacetime plot of $\braket{S^z}(x,t)$; middle row: $\braket{\mathcal
P_{\uparrow\uparrow}}(x,t)$, which projects a bond onto
$\ket{\uparrow\uparrow}\bra{\uparrow\uparrow}$; bottom row: $\braket{\mathcal
P_{\uparrow\uparrow\uparrow}}(x,t)$, which projects three adjacent sites onto
$\ket{\uparrow\uparrow\uparrow}\bra{\uparrow\uparrow\uparrow}$.}
\label{fig:three spin excitations}
\end{figure}

In all other cases $m_0>1$, string states $\ell\geq 2$ will contribute to the time
evolution of observables 
and in order to study their influence we have
carried out numerical computations using the TEBD
\cite{tebd}.
Results for $m_0=3$ (three neighbouring sites with spin up in the
initial state) are shown in Fig.~\ref{fig:three spin excitations}. 
As a function of the anisotropy $\Delta$ we observe three distinct
regimes, which are fully consistent with expectations from the
Bethe ansatz: (i) for small values of $\Delta$ we observe a single
wave front in $\langle S^z\rangle(x,t)$, propagating with the maximal
magnon velocity $v=J$ (the $m_0=1$ case discussed above looks quite similar).
(ii) At $\Delta=0.8$, a second, slower branch of propagating wave
packets emerges both in $\langle S^z\rangle(x,t)$ and in
$P_{\uparrow\uparrow}(x,t)$ \endnote{We note that these wave fronts, while
evident in Fig.~\ref{fig:three spin excitations}, are not easy to
discern in equal time or space slices due to the oscillatory nature of
the signal.}. Its propagation velocity is equal to the
maximal 2-string velocity. We have verified by direct evaluation of
(\ref{lehmann}) that the second front is associated with 2-strings.
Interestingly there is a threshold in $\Delta$ for observing this
phenomenon ($\Delta_{c}\approx\Delta_{0}=1/\sqrt{2}$), while
2-strings exist at any $\Delta\neq 0$. The 
reason is that the maximal 2-string velocity is 
$v_{\rm max,2}=J\sqrt{1-\Delta^2}$ for $0<\Delta<\Delta_0$ and
$v_{\rm max,2}=\frac{J}{2\Delta}$ for $\Delta_0<\Delta<1$.
On the other hand, the density of 
states for 2-strings is $\rho_2(v)=2\Delta/\sqrt{J^2-(2\Delta v)^2}$,
which acquires a singularity only if $\Delta>1/\sqrt{2}$. It is this
singularity which induces a clear signature of propagating 2-strings
in both $\langle S^z\rangle(x,t)$ and $P_{\uparrow\uparrow}(x,t)$. 
(iii) For interaction strengths above $\Delta_{c2}\approx 0.9$ we
observe an additional branch in $\langle S^z\rangle(x,t)$, 
$P_{\uparrow\uparrow}(x,t)$ and in
$P_{\uparrow\uparrow\uparrow}(x,t)$. This feature clearly arises from
propagating 3-strings and can be understood in complete analogy with
the 2-string case discussed above.

\begin{figure}
\includegraphics[width = 5.4cm,height=4.4cm]{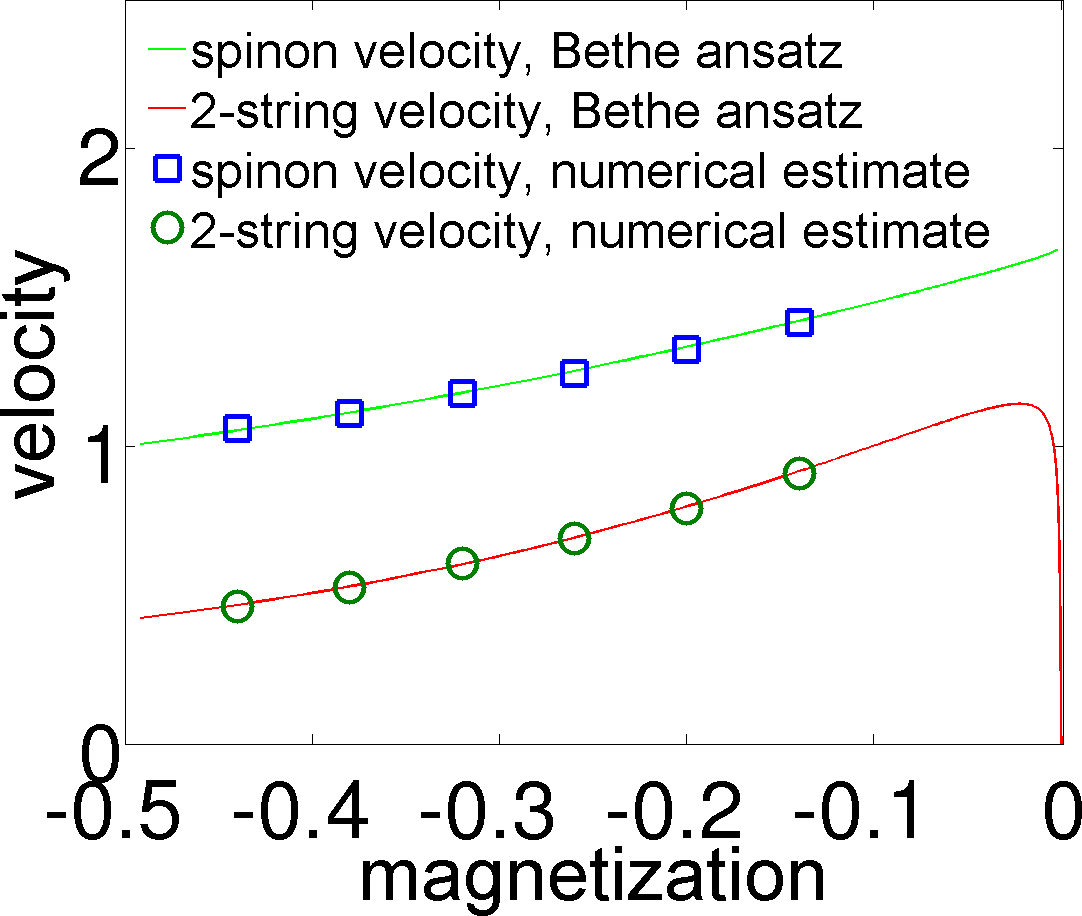}
\caption{
Propagation velocity of single-spinon and two-string branch as a function of total
magnetization per site
$m$ of the system at $\Delta = 1.2$. Green and red curves show single-spinon and 
two-string velocities as calculated from Bethe ansatz. Blue circles and squares are
numerically derived values from real time simulations. Errorbars are smaller than symbols.}
\label{fig:veloc}
\end{figure}

\emph{Results for Finite Magnetizations:} Here the bulk of our system
is in a strongly correlated quantum critical Luttinger liquid phase
and our quench protocol described above is closely related to the
X-ray edge singularity problem in a correlated host
\cite{affleckludwig}. However, the observables 
relevant to our case are different and cannot be described using
methods of boundary conformal field theory \cite{boundaryCFT}.
We computed the quenched ground state using the density matrix renormalization 
group algorithm \cite{DMRG} and the time evolution using the
TEBD with matrix dimensions up to 1200.
\begin{figure}
\begin{minipage}{0.2\textwidth}
$\braket{S^z}(x,t)$
\end{minipage}
\hspace{0.25cm}
\begin{minipage}{0.2\textwidth}
$\mathcal{P}_{\uparrow\uparrow}(x,t)$
\end{minipage}

  \includegraphics[width =8cm, height=8.7cm]{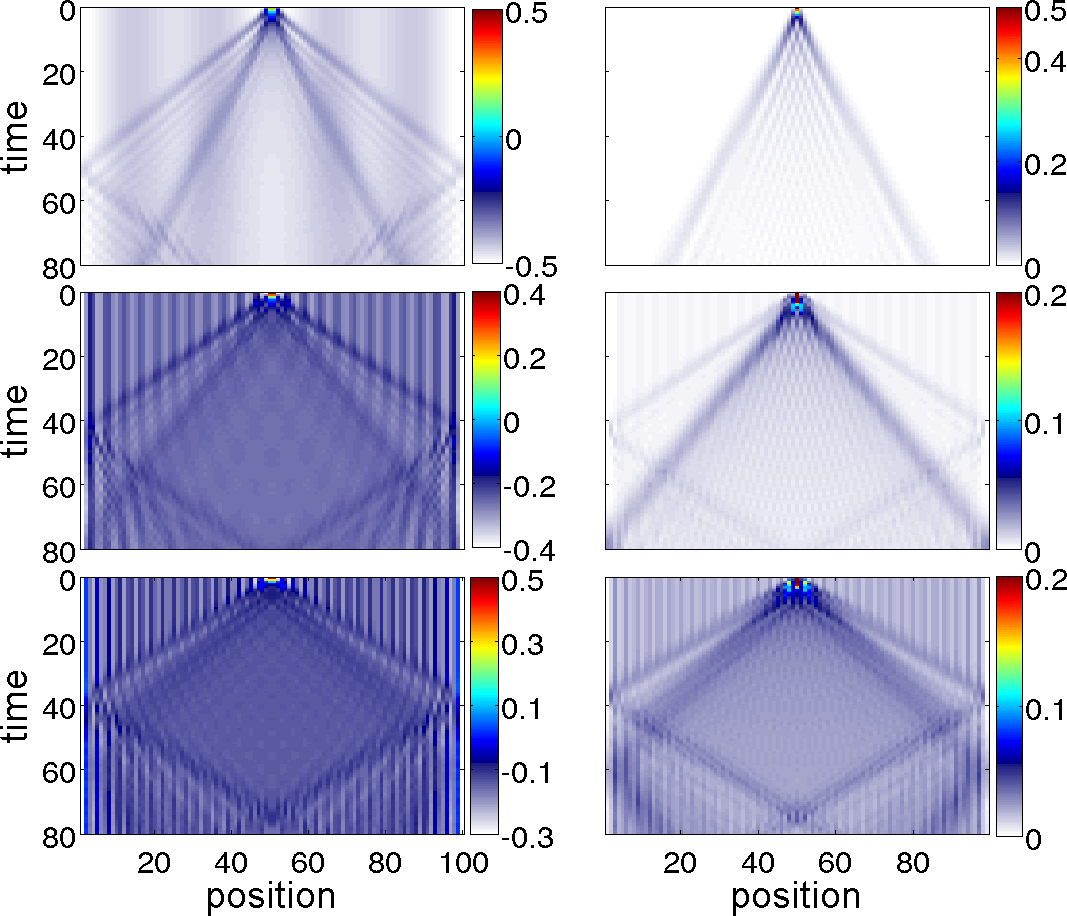}
\caption{Two-string propagation at finite magnetization per site
$m$ at $\Delta=1.2$, corresponding to the Luttinger liquid phase
of the model. From top to bottom, ${m}=-0.44$,
${m}=-0.26$ and ${m}=-0.14$. The initial state
at $t=0$ is the ground state of (\ref{eq:Heisenberg hamiltonian})
with an infinite magnetic field term at two sites in the center of
the chain (chain length $N$=100). At $t=0$, the field is switched off and the state is
evolved. The striped patterns visible in all plots are Friedel oscillations due to open boundary conditions.
}
\label{fig:two spin finite densities}
\end{figure}
In Fig.~\ref{fig:two spin finite densities} we present results for
$\Delta=1.2$ and three different magnetizations per site 
$m=({N_\uparrow-N_\downarrow})/2N=-0.44,-0.26,-0.14$, corresponding
to $N_\uparrow=6,24,36$ on a $N=100$ site chain. We note that this
corresponds to the Luttinger liquid phase of the Heisenberg model even
though $\Delta>1$. In all cases we observe two propagating wave fronts
(in each direction) in $\langle S^z\rangle(x,t)$. The results for
$P_{\uparrow\uparrow}(x,t)$ show that the slower front is associated
with excitations that favour neighbouring spin flips. 
In order to interpret these results we follow
our analysis of the spin polarized case. It is known from the Bethe
ansatz solution that the elementary excitations of the Heisenberg
chain at finite magnetization are gapless ``spinons'' as well as
gapped bound states associated with string solutions of the Bethe
ansatz equations (\ref{bae}). It is then tempting to associate the
faster/slower wave fronts with spinon and 2-string excitations
respectively, because, just like in the spin polarized case, the
latter induce an enhancement in the density of neighbouring spin flips
as a result of their bound nature. In order to substantiate this
expectation we have evaluated the maximal velocities of both spinon
and string excitations as functions of the magnetization per site.
In Fig.~\ref{fig:veloc} 
we present a comparison of these
velocities with the ones extracted from the TEBD results in
Fig.~\ref{fig:two spin finite densities}. We see that the results are
in excellent agreement.
\begin{figure}
 \includegraphics[width=5.9cm]{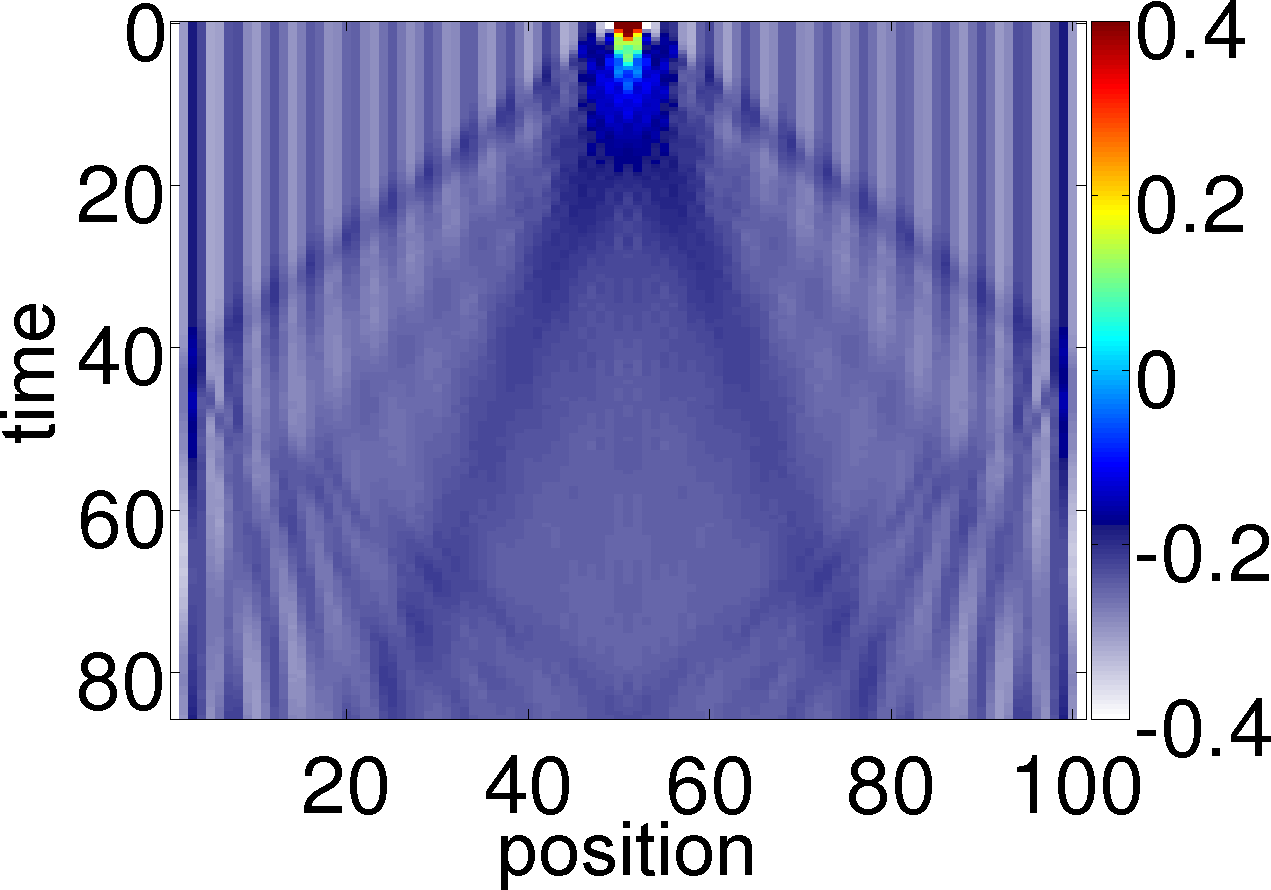}
\caption{Spacetime plot of $\braket{S^z}$ for a setup
similar to Fig.~\ref{fig:two spin finite densities} with $N$ = 101 at
total magnetization $m=-0.2525$, $\Delta=1.2$, with
three particles at the chain center at $t$=0.}
\label{fig:AF3}
\end{figure}

For magnetizations closer to zero the two-string
branch gets more and more washed out, because 
the momentum range of two-string
excitations diminishes and eventually vanishes as the magnetization
approaches zero \cite{strings}. 
In order to determine whether longer strings also
lead to easily recognizable features in observables after a
local quench we have analyzed the case $m_0=3$ for $\Delta=1.2$ and
magnetization per site $m=-0.2525$. Some results for 
$\langle S^z\rangle(x,t)$ are shown in 
Fig.~\ref{fig:AF3}. 
We can now identify three branches. The propagation
velocities extracted from the TEBD data are 
$v_1\approx 1.26\pm 0.02$, $v_2\approx 0.702\pm 0.025$ and $v_{3}
\approx 0.370\pm 0.02$ respectively. These values agree with the
maximal velocities of spinons, 2-strings and 3-strings calculated from
Bethe ansatz, which are $v_{\rm max}\approx 1.263$, $v_{\rm
max,2}\approx 0.705$ and $v_{\rm max,3}\approx 0.375$. 

\begin{figure}
\includegraphics[width=8cm]{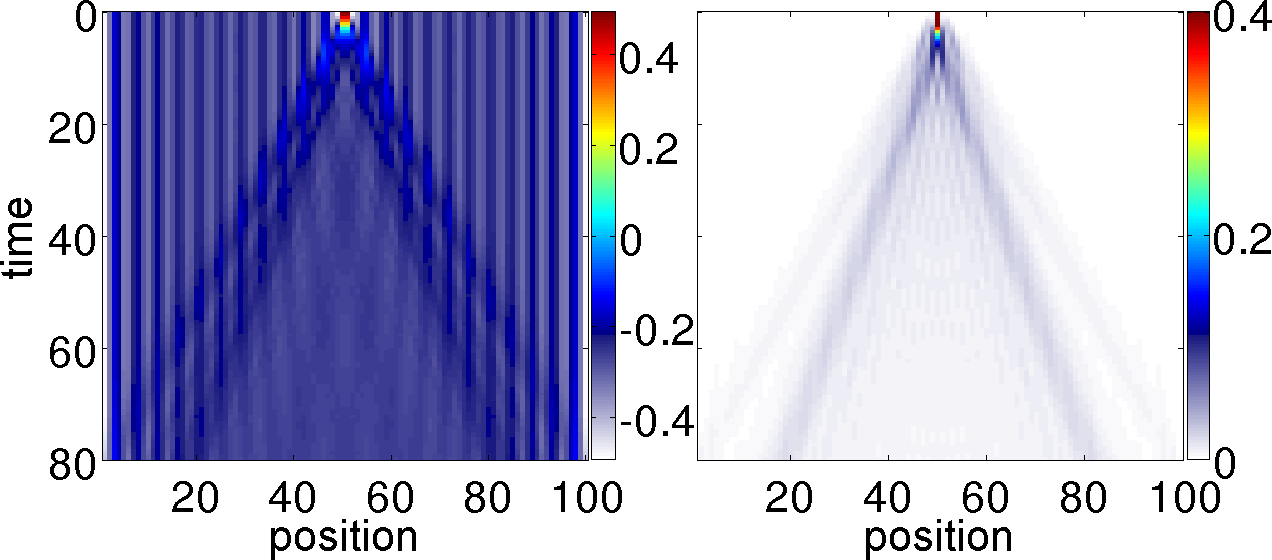}
\caption{Spacetime plot of $\braket{S^z}$ and $P_{\uparrow\uparrow}$
for $N$ = 100, $m_0=2$ at total magnetization $m=-0.26$, $\Delta=1.2$ and
an extra integrability breaking term $(J/10)\sum_j {\bf S}_j\cdot{\bf
  S}_{j+2}$ added to $H(\Delta,B_0)$. Bound state signatures are seen
to persist.}
\label{fig:nnn}
\end{figure}

\textit{Integrability breaking perturbations:}
In general, string states are not protected kinematically from decaying
into scattering states of spinons. Their stability is then a
consequence of integrability of the Heisenberg chain and an important
question is, whether signatures of bound states survive when
integrability breaking perturbations are taken into account. In order
to address this issue we have considered two types of perturbation:
(i) a next-nearest neighbour interaction and (ii) a spatially varying
magnetic field term $\gamma\sum_{j=1}^N (j-\frac{N}{2})^2S^z_j$, which
would model an optical trap in certain realizations of
(\ref{eq:Heisenberg hamiltonian}) based on cold fermionic atoms. In
both cases we observe signatures of bound states, indicating that they
survive in the form of resonances. We show results for case (i) in Fig.~\ref{fig:nnn}.

\textit{Conclusions:}
We have studied local quantum quenches in the antiferromagnetic
spin-1/2 Heisenberg XXZ chain. We observed that above certain
thresholds in the interaction strength $\Delta$, local observables
exhibit prominent signatures associated with linearly propagating gapped bound
states. Given the difficulty in observing these bound states in
scattering experiments on quantum magnets \cite{Pereira_2008,structurefactor}
we propose that non-equilibrium setups of the kind
considered here are an ideal setting for observing them and probing
their properties. Heisenberg spin chains can, e.g., be realized
experimentally in crystals and systems of cold atoms in optical lattices 
with time and space resolved dynamics \cite{Crystal,Resolved_Dynamics_optical_lattice}.
Recent work has focussed on AC driven optical lattices \cite{experiment} and
two-component Bose mixtures \cite{2comp}. The kind of local
perturbation characterizing our initial state could be induced by a
focussed laser beam.

\textit{Acknowledgements} 
We thank P. Calabrese, J. Cardy 
and A. Silva 
for helpful comments and
discussions. This work was supported by the Austrian Science Fund
(FWF) within the SFB ViCoM (F41) and by the EPSRC under grant
EP/I032487/1. FHLE and HGE thank the KITP for hospitality. This
research was supported in part by the NSF under grant No. NSF PHY05-51164.



%
%

\end{document}